\documentclass[prl,twocolumn]{revtex4}
\usepackage[ansinew]{inputenc} 
\usepackage{graphicx} 
\usepackage{color} 
\usepackage{amsmath} 
\usepackage{amssymb} 
\usepackage{units} 

\newcommand{\bk}{{\bf k}}
\newcommand{\be}{\begin{equation}}
\newcommand{\ee}{\end{equation}}
\renewcommand{\r}{{\bf r}}

\newcommand{\R}{{\bf R}}

\begin{document}

\title{Conservation of energy in coherent backscattering of light}

\author{S. Fiebig$^{(1)}$, C.M.~Aegerter$^{(1)}$, W. B\"uhrer$^{(1)}$, M. St\"orzer$^{(1)}$,
E. Akkermans$^{(2)}$, G. Montambaux$^{(3)}$, and G. Maret$^{(1)}$}
\affiliation{$^{(1)}$ Fachbereich Physik,
University of Konstanz, Box M621, 78457 Konstanz, Germany \\
$^{(2)}$ Department of Physics, Technion Israel Institute of
Technology, 32000 Haifa, Israel \\
$^{(3)}$ Laboratoire de Physique des Solides, CNRS UMR 8502,
Universit\'e Paris-Sud, 91405 Orsay, France }

\date{\today}

\begin{abstract}
Although conservation of energy is fundamental in physics, its
principles seem to be violated in the field of wave propagation in
turbid media by the energy enhancement of the coherent
backscattering cone. In this letter we present experimental data
which show that the energy enhancement of the cone is balanced by
an energy cutback at all scattering angles.  Moreover,
we give a complete theoretical description, which is in good
agreement with these data. The additional terms needed to enforce
energy conservation in this description result from an
interference effect between incident and multiply scattered waves,
which is reminiscent of the optical theorem in single scattering.
\end{abstract}

\maketitle


Conservation of energy is one of the most fundamental principles
in physics. Not only in mechanics, but also in thermodynamics it
rests on fundamental symmetries, which so far have not been
violated. There have however been instances, where novel effects
gave rise to suspicions of non-conservation of energy as in the
distribution of positron energies in beta decay. This was only
explained by the introduction of the neutrino in Fermi's
four-point theory \cite{fermi}. Another instance which seemingly
contradicts the conservation of energy is the coherent
backscattering cone which appears when waves propagate in turbid
media. In that case, a twofold enhancement of the backscattered
intensity with respect to Lambert's law for diffuse scattering is
observed. This enhancement decays over an angular scale of $(k
l^*)^{-1}$, where $k$ is the wavenumber of the wave and $l^*$ is
the transport mean free path of the medium \cite{akkermans}. Due
to its fundamental nature, it can be observed in areas as
different as solid state physics
\cite{Kaveh_(PRL_57_2049_(1986))}, soft matter physics
\cite{van_Albada_(PRL_55_2696_(1985))}, astro- \cite{moon} and
geophysics \cite{seismiccone}, and with various kinds of waves
like sound waves \cite{acoustic}, microwaves \cite{microwaves}
and, as in the experiments presented in this letter, visible light
\cite{van_Albada_(PRL_55_2696_(1985)),ishimaru}.

The additional energy contribution of the coherent backscattering
cone to the intensity can however not be explained by a
corresponding reduction in transmission at surfaces not considered
in the experiment. Rather, the backscattering cone is also
observed from samples which can theoretically as well as
experimentally be treated as filling an infinite half-space,
meaning that the waves can reach no other surface than the one
considered \cite{akkermans}. Moreover, different polarization
channels also show a backscattering cone \cite{akkermansjphys},
such that the energy in the cone cannot be obtained from another
polarization channel.

The origin of the backscattering cone lies in the interference of
waves propagating along reciprocal paths. This interference can only
spatially re-distribute the backscattered energy. Thus the energy
enhancement at small angles has to be accompanied by a corresponding
energy cutback to ensure conservation of energy. However, such an
energy cutback has so far not been observed experimentally.
Moreover, it is not described by the prevailing theories
\cite{akkermans,akkermansjphys,vandermark}. This can be problematic
as the scaling of the width of the backscattering cone with $k l^*$
is commonly used to characterize multiple scattering materials. In
particular in turbid samples, when the cone becomes very broad,
there has to be a sizeable correction if conservation of energy is
to hold. This is of great importance in the study of Anderson
localization of light \cite{Anderson_(PhysRev_109_1492_(1958))},
where a reliable knowledge of the parameter $kl^*$ is needed to
characterize the phase transition from diffusive transport to an
insulating state \cite{Aegerter_(EurophysLett_75_(2006)),locprl}.

In this letter, we present measurements of coherent
backscattering, where the incoherent background is determined on
an absolute scale. With this we are able to show that there is a
reduction in backscattering intensity at all angles compensating
for the enhancement in the back-direction. A theoretical
description which fits these data shows that the reduction in
enhancement results from a new interference effect between the
incident and multiply scattered waves. This is analogous to the
shadow term, which accounts for flux conservation in the optical
theorem \cite{born}. In multiple scattering, the terms needed to
ensure energy conservation correspond to the so-called Hikami box
or quantum crossing \cite{hikami,akkermansbook}. The present
experiment constitutes a direct observation of a scattering
process described by a Hikami box, which plays a central role in
quantum mesoscopic physics \cite{akkermansbook}.

Our main setup to study the angular distribution of the
backscattered light consists of 256 photosensitive diodes attached
to a semi-circular arc with a diameter of $\unit[1.2]{m}$, in the
center of which the sample is located \cite{peter}. In this way,
we can detect light over a range of $-60^\circ < \theta <
85^\circ$ with a resolution of $0.14^\circ$ for $|\theta| <
10^\circ$, $\sim 1^\circ$ for $10^\circ < |\theta| < 60^\circ$ and
$\sim 3^\circ$ for $\theta > 60^\circ$. For the illumination a
continuous wave dye laser with a wavelength of $\unit[590]{nm}$ is
used. The measurements are done using circularly polarized light
in order to reduce the influence of single scattering. To average
over random speckle patterns, the samples are rotated. As the very
tip of the cone at $\theta \simeq 0$ can not be resolved with this
setup, the central part of the backscattering cone, $|\theta| <
3^\circ$, is measured separately using a beam splitter and a
charge-coupled device (CCD) camera to a resolution of $0.01^\circ$
\cite{peter}.

In order to calibrate the photodiodes, we have used a block of
teflon as a reference sample. Teflon has a transport mean free path
of $\simeq \unit[300]{\mu m}$ and hence the backscattering cone of
teflon at a wavelength of $\unit[590]{nm}$ has a FWHM of about
$0.02^\circ$, which is much narrower than the angular resolution of
the wide-angle setup. This implies that with the wide-angle setup
teflon can be considered as a purely incoherent signal, which is
properly normalized given by $\mu (\gamma + \mu /(\mu+1))$
\cite{akkermansbook}, where $\mu = \cos \theta$ and $\gamma l^*$
describes the distance over which the diffuse intensity enters the
sample. These diode signals are measured at several different
incident laser powers, which are determined independently with a
calibrated power-meter. Interpolation of the measured data then
yields a calibration function for each photodiode \cite{peter}.
\begin{figure}
\begin{center}
\includegraphics[width=\columnwidth]{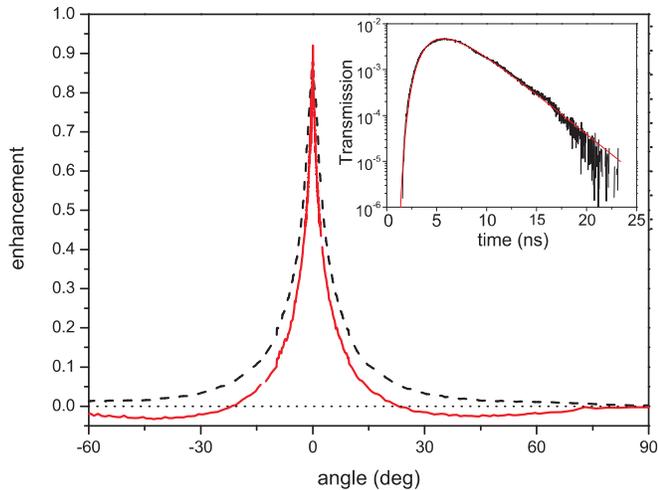}
\end{center}
\caption{\label{loss-lossless} Backscattering cone of R700
evaluated ignoring (dashed curve) and taking into account (full
red curve) absorption in the reference sample. The former is
positive for all angles, resulting in an uncompensated energy
enhancement of the cone. For the latter, energy enhancement and
energy cutback are balanced. This is quantified by the half-space
integral of the intensity $I = -0.005(7)$.
The inset shows a time of flight measurement of the teflon
reference, which follows diffusion theory with an absorption time
of 3.3 ns.}
\end{figure}
To be able to determine the intensity of the backscattered cone on
an absolute scale, the incoherent background needs to be known.
Since the cone of teflon cannot be resolved in the wide-angle
setup, the teflon reference measurements can be considered to
describe the incoherent background of the TiO$_2$ samples. In
doing so however, one neglects the different albedo of teflon with
respect to the sample. The proper background for the sample,
$\alpha_{\text{inc,samp}}$ is therefore given by that of the
reference, $\alpha_{\text{inc,ref}}$ multiplied by the ratio of
the albedos of sample and reference, $A_{\text{samp}}
/A_{\text{ref}}$.

For an estimation of the albedos to a level of better than one
percent, one needs to take into account losses at the
sample/reference boundaries, as well as losses due to absorption.
Up to now these losses have not been fully taken into account in
the evaluation of the backscattering cones
\cite{wiersmarsi,peter}.

The loss due to leakage is estimated by comparing the diffuse
energy in an infinite half-space at a certain time with the amount
of energy that is left in a volume-cutout of the infinite
half-space corresponding to the size of the reference
\cite{vandermark}. This will neglect edge effects, but since we
are only dealing with the very tails of the diffusive cloud, this
will be accurate to the desired level. For this estimate, the
diffusion coefficient of the sample/reference needs to be known.
The loss factor for absorption can be calculated from the integral
over the path length distribution $P(D,\tau,t)$, where $D$ is the
diffusion coefficient and $\tau$ is the absorption time.

Both quantities needed to calculate the albedo, $D$ and $\tau$ can
be determined with a time resolved transmission experiment
\cite{watson,locprl}, see the inset of Fig. \ref{loss-lossless}.
Such a time of flight experiment directly gives the path length
distribution inside the sample, which is a function only of $D$
and $\tau$. In our experiment, the same dye laser is used as in
the backscattering experiments, thus making a possible wavelength
dependence of $D$ and $\tau$ irrelevant. We thus calculate the
ratio of the albedos using the absorption length $L_a = \sqrt{D
\tau}$ with respect to the reference and hence the absolute level
of the incoherent background for all samples. Subtracting this
background then directly gives a proper measure of the backscatter
enhancement.


\begin{figure}
\begin{center}
\includegraphics[width=\columnwidth]{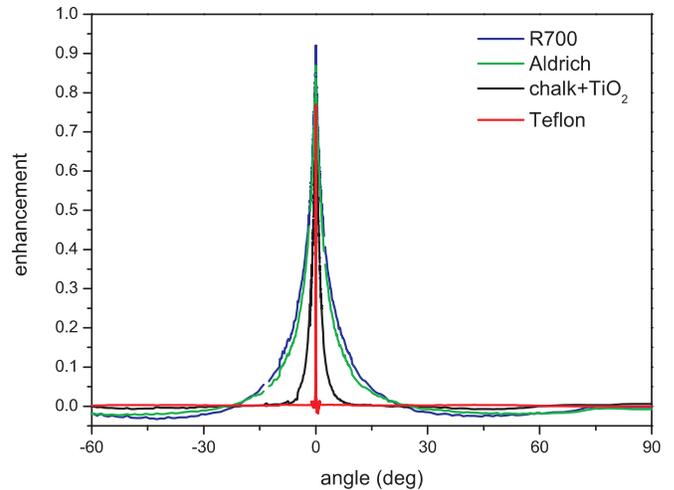}
\end{center}
\caption{\label{enhancement-angle} Measurements of backscattering
cones for different samples. As the cone width increases, more
energy needs to be compensated. Thus the most turbid samples (R700
and Aldrich) lead to a noticeable energy cutback at angles around
$45^\circ$. The amplitude of this cutback is reduced with decreasing
turbidity, but stays positioned around $45^\circ$. This indicates
that the cutback is due to effects occurring at a fixed length scale
close to $\lambda$.}
\end{figure}
To represent cones with a wide variety of cone widths, we have used
samples of ground TiO$_2$ particles in its rutile structure with
different particle diameters (R700: 245 nm and Aldrich: 540 nm), a
mixture of TiO$_2$ (R700) and ground chalk in a weight ratio of one
to five, as well as solid teflon. The TiO$_2$ particles are
commercially available as pigment for white paint \cite{locprl}.

As can be seen in Figs. \ref{loss-lossless} and
\ref{enhancement-angle}, a cutback of the backscattered energy is
indeed observed when taking into account the different loss
factors of reference and sample. This is most significant for
samples with very wide cones like R700 and Aldrich with
$\text{FWHM}\approx 3.8^\circ$ and $3.4^\circ$ respectively, where
the enhancement is noticeably below zero for a range of $50^\circ$
on both sides of the cone, as shown in
Fig.~\ref{enhancement-angle}. The TiO$_2$-chalk-mixture with
$\text{FWHM}\approx 1.5^\circ$ still shows a slight lowering of
the enhancement, while the enhancement for teflon with
$\text{FWHM}\approx 0.013^\circ$ is essentially zero away from the
cone. Note that unlike the coherent backscattering peak, the
energy cutback is not characterized by a specific angle but it is
rather spread out over the whole angular range. Furthermore, these
energy cutbacks do compensate the energy enhancements of the
cones, as the integral of the enhancement $\alpha(\theta)$ over
the backscattering half-space $I = \int
\alpha(\theta) \sin{\theta} \; d\theta$ is zero for all
investigated samples within the margins of error. This shows that
a determination of the absolute intensity scale is crucial for the
correct observation of the backscattering cone. For such a
determination, the different loss factors of reference and sample
had to be accounted for.


Theoretically, the coherent backscattering cone has
been described in great detail \cite{akkermansbook}. In the
geometry of a semi-infinite medium of section $S$ (see Fig. \ref{overlap}a),
this description makes use of the following well-known expression
for the coherent albedo $\alpha_c ^A$,
\begin{equation}
\alpha_c ^A = {c \over 4 \pi S l^{*2}} \int d \r_1 d \r_2 H^A
(\r_1 , \r_2 ) P( \r_1 , \r_2 ) \label{albedoA} \ee where $H^A
(\r_1 , \r_2 ) = e^{- {\mu + 1 \over \mu} {z_1 + z_2 \over 2 l^*}}
e^{i (\bk + \bk') \cdot (\r_1 - \r_2)}$ for an incident plane wave
normal to the interface. $P(\r_1, \r_2 )$ is the probability of
having a multiple scattering path starting at $\r_1 $ and ending
at $\r_2$. The first factor in $H^A$ describes the attenuation of
an incident plane wave over a distance of the order of the elastic
mean free path $l^*$. The second factor in $H^A$ accounts for
interference and leads to the enhancement with an angular width of
order $1 / kl^*$.

The interference term $H^A$ is the product of four amplitudes
describing the two incoming and the two outgoing plane waves. It
is known as a quantum crossing and it is at the origin of coherent
effects in quantum mesoscopic physics such as weak localization,
universal conductance fluctuations and eventually it leads to the
localization transition. Energy (or number of particles)
conservation imposes  constraints on the quantum crossings. It is
well-known \cite{hikami} that in order to fulfill this constraint,
two other contributions $H^{B,C} (\r_1 , \r_2 )$ must be added to
$H^A$, which mix the in- and outgoing wave vectors. Energy
conservation thus imposes that $ \int d \R \, \ (H^A +H^B + H^C )
=0$, where $\R = \r_1 - \r_2$.

\begin{figure}
\begin{center}
\includegraphics[width=\columnwidth]{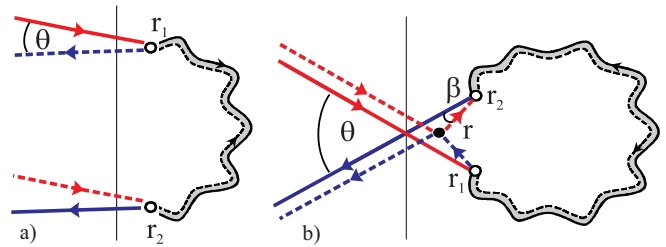}
\end{center}\caption{\label{overlap} Wave configurations of the
contributions $H^A$ and $H^{B,C}$ to coherent backscattering.
$H^A$ (a) describes interference between time reversed amplitudes
(full and dashed lines) and gives the classical cone shape
\cite{akkermans}. When $\r_1$ and $\r_2$ are within a transverse
distance of $\lambda$, the time-reversed loops have to be
considered as closed and the amplitudes are becoming coupled. This
is described by $H^{B,C}$ (b), as an interference effect between
the incident plane wave and the attenuated spherical wave
traveling between the points $\r_{1,2}$ and a newly introduced
scatterer $\r$ at an angle $\beta$ (see text). }
\end{figure}

A complete description of coherent backscattering must therefore
include these additional contributions $H^{B}$ and $H^{C}$, which
are equal. The physical basis of these contributions lies in a
coupling of the light fields at the first and the last scatterer
($\r_1$ and $\r_2$), when they are within a volume of order
$\lambda^2 l^*$. This coupling originates in an interference of
the incoming plane wave with the multiply scattered spherical wave
and is described by introducing an additional scattering event
located in $\r$ (see Fig.\ref{overlap}b). This is reflected by the
short range behavior of $H^B(\r_1,\r_2)$ and the additional
contribution results from almost closed diffusive trajectories.
Consequently, this interference is not restricted to small angles
$\theta$ as is the case for $\alpha_c^A$.
For the case of an incident wave normal to the interface, the
contribution $\alpha_c ^B$ can be written as:

\be \alpha_c ^B \simeq {c \over S l^{*3}} \int d \r  P( \r , \r )
e^{- {\mu + 1 \over \mu} {z \over  l^*}} h^2, \label{albedoB} \ee
where \be h  \simeq - \int d \r'  \, \ e^{i \bk' \cdot \r'} {e^{i
k r'} \over 4 \pi r'} e^{- r'/ 2l^*} \simeq {i l^* \over 2 k}
\label{imu} \ee is calculated to leading order in $(kl^*)^{-1}$.
It is interesting to note the similarity between $h$ and the
shadow term which occurs in the optical theorem and ensures flux
conservation for single elastic scattering. Here, the shadow terms
$h$ describe the interference between the incident and the
multiply scattered waves. The integral in (\ref{albedoB}) can be
solved approximately to give a correction $\alpha^{B+C}_c \simeq -
1.15 (kl^*)^{-2} \mu/(\mu+1)$. Thus the correction is of order $-
(kl^*)^{-2}$. Noting that the angular integral of $\alpha_c^A
\simeq (kl^*)^{-2}$, we indeed retrieve the energy conservation
condition of quantum crossing, namely that the integral over
$\alpha_c ^{A+B+C}$ is zero, as it should be.

The expression for $\alpha^{B+C}_c$ shows that the interference
effect in (\ref{albedoB}) is twofold. First, $h$ is
purely imaginary so that $h^2$ is negative resulting in a depletion
of the coherent albedo proportional to $(kl^*)^{-2}$. Secondly,
this interference term does not contribute at a specific angular
value but it is rather spread out over the whole angular range.

A fit of the total coherent albedo $\alpha_c ^{A+B+C}$ to our data
is shown in Fig. \ref{measurement-theory} by the dashed line.
\begin{figure}
\begin{center}
\includegraphics[width=\columnwidth]{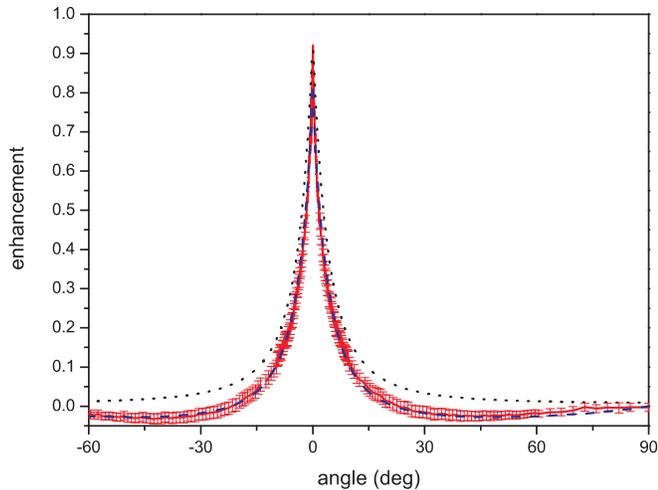}
\end{center}
\caption{\label{measurement-theory} Comparison of the
backscattering cone of R700 with corrected and uncorrected
enhancement. The agreement between the measured data and the fit
of the corrected enhancement $\alpha^{A+B+C}_c$
(dashed) is perfect within the errors. The uncorrected enhancement
 $\alpha^A_c$ (dotted) calculated with the value of
$k l^*$ obtained by the fit of the data with
$\alpha^{A+B+C}_c$ describes the cone itself quite well, but
shows significant deviations in the area of the energy cutback.}
\end{figure}
The dotted line shows only $\alpha^A_c$, where the
same value of $kl^*$ has been used as in $\alpha^{A+B+C}_c$.
Comparing this description with the uncorrected data in
Fig.~\ref{loss-lossless} shows that the $kl^*$ values
determined with those data are close to their real values.

In order to properly determine the turbidity of the sample, the
value of $kl^*$ obtained from (\ref{albedoA}) still has to be
corrected for internal reflections \cite{Zhu_(PRA_44_3948_(1991))},
which lead to a broader distribution of light paths, and  thus to a
narrower cone. Furthermore, the ratio between the transport and
elastic mean free paths will influence the pre-factor in the
correction but is assumed to be close to unity.



In conclusion, we have shown experimentally that coherent
backscattering does fulfill conservation of energy. To this purpose,
the losses of the reference sample had to be quantified via a time
of flight measurement to ensure an absolute energy calibration of
the setup. If the loss of the reference differs significantly from
the loss of the sample, this leads to different positions of the
incoherent background in spite of equal incident laser energies.

Furthermore, we have provided a complete theoretical description
of coherent backscattering based on the calculation of the
 three terms $H^{A,B,C}$ that contribute to the
Hikami box or quantum crossing. $H^A$ describes the steep angular
variation around backscattering and $H^{B,C}$ take into account
crossed diagrams dressed by a scattering impurity. Such an impurity
provides an additional interference between incoming and multiply
scattered waves at short distances.  Since the cone is basically the
Fourier transform of this intensity this leads to a broadly
distributed energy cutback \cite{akkermansjphys}. This improved
description of the cone-shape for extremely turbid samples also
allows a reliable determination of $kl^*$ in such samples. A
comparison of these result with those obtained previously on the
same samples show however that the values of $kl^*$ thus determined
change only very little. This implies that the systematic dependence
of deviations from classical diffusion is indeed as it was
previously determined \cite{Aegerter_(EurophysLett_75_(2006))}.


\begin{acknowledgments}
This work was supported by the Deutsche Forschungsgemeinschaft,
the International Research and Training Group "Soft Condensed
Matter of Model Systems", the Center for Applied Photonics
(CAP) at the University of Konstanz, the Israel Academy of
Sciences and by the Fund for Promotion of Research at the
Technion. Furthermore, we would like to
thank Aldrich and DuPont chemicals for providing samples used in
this study. We also appreciate very much Peter Gross' help in
building the backscattering cone setup.
\end{acknowledgments}



\begin{thebibliography}{99}

\bibitem{fermi} E. Fermi, Z. Phys. {\bf 88}, 161 (1934).
\bibitem{akkermans}
E. Akkermans, P.E. Wolf, and R. Maynard, Phys.\ Rev.\ Lett.\
\textbf{56}, 1471 (1986).
\bibitem{Kaveh_(PRL_57_2049_(1986))}
M. Kaveh {\em et al.}, Phys. Rev. Lett. \textbf{57}, 2049 (1986).
\bibitem{van_Albada_(PRL_55_2696_(1985))}
M.P. van Albada, and A. Lagendijk, Phys.\ Rev.\ Lett.\
\textbf{55}, 2692 (1985); P.E. Wolf, and G. Maret, {\em ibid} {\bf
55}, 2696 (1985).
\bibitem{moon} B.W. Hapke, R.M. Nelson and W.D. Smythe, Science {\bf 260}, 509 (1993).
\bibitem{seismiccone} E. Larose {\em et al.}, Phys. Rev. Lett. {\bf 93},
048501 (2004).
\bibitem{acoustic} G. Bayer and T. Niederdr\"ank, Phys. Rev. Lett. {\bf 70}, 3884 (1993).
\bibitem{microwaves} R. Dalichaouch {\em et al.}, Nature (London) {\bf 354}, 53
(1991).
\bibitem{ishimaru} Y. Kuga and A. Ishimaru, J. Opt. Soc. Am. A, {\bf 1}, 831 (1984).
\bibitem{akkermansjphys} E. Akkermans {\em et al.}, J. de Physique (France), {\bf 49}, 77 (1988).
\bibitem{vandermark} M.B. van der Mark {\em et al.}, Phys. Rev. B {\bf
37}, 3575 (1988).
\bibitem{Anderson_(PhysRev_109_1492_(1958))}
P.W. Anderson, Phys. Rev \textbf{109}, 1492 (1958).
\bibitem{Aegerter_(EurophysLett_75_(2006))}
C.M. Aegerter {\em et al.}, Europhys. Lett. {\bf 75}, 562 (2006).
\bibitem{locprl} M. St\"orzer {\em et al.}, Phys. Rev. Lett. {\bf 96}, 063904 (2006).
\bibitem{born} M.~Born and E.~Wolf, {\em Principles of Optics}, Oxford University Press (1980).
\bibitem{hikami} S. Hikami, Phys. Rev. B {\bf 24}, 2671 (1981).
\bibitem{akkermansbook} E. Akkermans and G. Montambaux, {\em Mesoscopic Physics of electrons and
photons}, Cambridge University Press, (2007).
\bibitem{peter} P. Gross {\em et al.},  Rev. Sci. Instr. {\bf 78}, 033105 (2007).
\bibitem{wiersmarsi} D.S.~Wiersma, M.P. van Albada, and A. Lagendijk, Rev. Sci. Instr. \textbf{66}, 5473 (1995).
\bibitem{watson} G.H. Watson {\em et al.}
, Phys. Rev. Lett. {\bf 58}, 945 (1987).
\bibitem{Zhu_(PRA_44_3948_(1991))}
J.X. Zhu {\em et al.},
Phys. Rev. A \textbf{44}, 3948 (1991).

\end{thebibliography}
\end{document}